\newcommand{\Tr}{\mathrm{Tr}}
\newcommand{\atanh}{\mathrm{atanh}}
\newcommand{\argmin}{\mathrm{argmin}}

\documentclass[conference]{IEEEtran}
\ifCLASSINFOpdf
\else
   \usepackage[dvips]{graphicx}
\fi
\begin{document}
%\title{Bare Demo of IEEEtran.cls for Conferences}
\title{Correcting beliefs in the mean-field and Bethe approximations using linear response}

\author{\IEEEauthorblockN{Jack Raymond}
\IEEEauthorblockA{Dipartimento di Fisica, Universit\`a La Sapienza, \\
Piazzale Aldo Moro 5, I-00185 Roma, Italy \\
Email: jack.raymond@physics.org}
\and
\IEEEauthorblockN{Federico Ricci-Tersenghi}
\IEEEauthorblockA{1. Dipartimento di Fisica, Universit\`a La Sapienza, \\
Piazzale Aldo Moro 5, I-00185 Roma, Italy \\
2. INFN--Sezione di Roma 1, and CNR--IPCF, UOS di Roma \\
Email: federico.ricci@roma1.infn.it}
%\and
%\IEEEauthorblockN{James Kirk\\ and Montgomery Scott}
%\IEEEauthorblockA{Starfleet Academy\\
%San Francisco, California 96678-2391\\
%Fax: (888) 555--1212}
}


%Telephone: (800) 555--1212, Fax: (888) 555--1212}


\maketitle

\begin{abstract}
%\boldmath
Approximating marginals of a graphical model is one of the fundamental problems in the theory of networks.
In a recent paper a method was shown to construct a variational free energy such that the linear response estimates, and maximum entropy estimates (for beliefs) are in agreement, with implications for direct and inverse Ising problems~\cite{Raymond:MFM}.
In this paper we demonstrate an extension of that method, incorporating new information from the response matrix, and we recover the adaptive-TAP equations as the first order approximation~\cite{Opper:TAPMpre}.
The method is flexible with respect to applications of the cluster variational method, special cases of this method include Naive Mean Field (NMF) and Bethe. 
We demonstrate that the new framework improves estimation of marginals by orders of magnitude over standard implementations in the weak coupling limit. 
Beyond the weakly coupled regime we show there is an improvement in a model where the NMF and Bethe approximations are known to be poor for reasons of frustration and short loops.
\end{abstract}


% \fi
\IEEEpeerreviewmaketitle

\section{Introduction}
%This demo file is intended to serve as a ``starter file''
%for IEEE conference papers produced under \LaTeX\ using
%IEEEtran.cls version 1.7 and later.
%I wish you the best of success.

Bethe and Naive Mean Field (NMF) are two of the most used variational methods, and can be considered special cases of the cluster variational method (CVM)~\cite{An:CVM,Pelizzola:CVM,Wainwright:GME}. Fast and provably convergent methods are known for the minimization of the CVM free energy~\cite{Yuille:cccpalgorithms}, and systematic expansion methods about minima have been shown~\cite{Chertkov:LC,Zhou:RGPF}. Generalizations and convex approximations to CVM have allowed for the development of fast and secure inference methods~\cite{yedidia,Weiss:MELP}. Linear response (LR) is often applied to the variational framework to determine accurately pair correlation estimates~\cite{Opper:TAPMpre,Welling:LRA,Kappen:BML,Montanari:CLC}.

In studying graphical models one is often interested in the estimation of probability distributions over local variables, and pair correlations; these can for example form the basis for decimation methods or moment matching algorithms~\cite{Montanari:SCS,RicciTersenghi:CMD,Wainwright:GME}. In this paper we develop an extension of the standard method for estimating marginal probabilities and pair-correlations in CVM, such that the variational parameters (beliefs) are self-consistent with the LR estimates. Our model improves over standard implementations on arbitrary graphs if couplings are weak. From the NMF framework we recover the adaptive-TAP equations for pairwise spin models~\cite{Opper:TAPMpre}. From the Bethe approximation, we recover the Sessak-Monasson expression for correlation estimation from a variational framework~\cite{Sessak:SME}. We apply the method to the homogeneous triangular lattice model: Bethe and NMF are known to perform poorly on this model due to the presence of short loops and frustration. For brevity we will focus only on binary variables (spins), pairwise couplings and two simple CVM approximations; but the principle is more general and might be combined with some of the aforementioned expansions and algorithmic methods.
 
%\hfill January 25, 2013

\subsection{Cluster variational method and linear response}
%Subsection text here.

Consider a model defined over $N$ spin variables $\{\sigma_i=\pm 1\}$, interacting through a symmetric coupling matrix $J$ and with external fields $H$. We can write the Hamilonian (Cost function);
\begin{equation}
\mathcal{H}(\sigma) = - \sum_{s \in I} J_{s}\prod_{i \in s} \sigma_i - \sum_i H_i \sigma_i \label{eq:Ham}\;.
\end{equation}
$s$ are subsets, and $I$ defines the set of non-zero couplings, for the pairwise model $I=\{(i,j) : J_{ij}\neq 0 \}$.
The exact free energy (cumulant generating function) of this model is
\begin{equation}
F = - \log \Tr [\exp(-{\mathcal H}(\sigma))]
\end{equation}	
$\Tr[\cdot]$ is a sum over the states in the enclosed expression. From the free energy one can calculate quantities of interest such as the magnetizations and pair connected correlations by linear response. Consider the connected correlation over a subset of variables $s$, it would be calculated as
\begin{equation}
  \chi_{s} = - \prod_{i \in s}\left[\frac{\partial}{\partial H_i}\right] F\;.\label{eq:linearresponse}
\end{equation} 
Naturally the application of linear response is not restricted to the exact free energy, using an approximate free energy one obtains pseudo statistics.
In either scenario one can construct the marginal (pseudo) probabilities, for example
\begin{equation}
  b^{LR}_{i} = \frac{1+\chi_i \sigma_i}{2}\;;\;b^{LR}_{ij}=b^{LR}_{i}b^{LR}_{j} + \frac{\chi_{ij}\sigma_i\sigma_j}{4} \label{eq:linearresponsemarginals}\;.
\end{equation}
Unfortunately the evaluation of the free energy is computationally intractable for many networks of interest, as such we cannot construct the marginals (\ref{eq:linearresponsemarginals}) by method (\ref{eq:linearresponse}) without approximations to the free energy.

The CVM provides such an approximation, based on a weighted sum of local entropy and energy contributions. The variational entropy and energy approximations are defined 
\begin{eqnarray}
	\!S(b)\! &=&\!\! - \sum_R c_R \Tr [b_R(\sigma_R) \log b_{R}(\sigma_R)] \label{eq:entropyCVM}\;;\\
	\!E(b)\! &=& \!\!- \sum_{s \in I} \Tr [J_s b_{s}(\sigma_s) \prod_{i \in s}\sigma_i] \!-\! \sum_i H_i \Tr [b_i(\sigma_i)\sigma_i]\;;\label{eq:energyCVM}
\end{eqnarray}
$R$ are regions (subsets of variables), $b_R$ are beliefs (approximate marginal probabilities) over the set of variables $\sigma_R$ (we henceforth omit the arguments $b_R=b_R(\sigma_R)$ for brevity), and $c_R$ are counting numbers. 

The theory of junction trees provides a justification for the selection of regions and counting numbers~\cite{Wainwright:GME}, if one selects regions according to a junction tree then one can recover the exact free energy by a constrained minimization
\begin{equation}
	F_{CVM} = \min_b\{E(b) - S(b) \} \;,
\end{equation}
subject to $0\leq b_R\leq 1$ and constraints
\begin{eqnarray}
0&=&\Tr [b_R]\;, \qquad \forall R \;;\label{eq:constraint}\\
0&=&\Tr_{\setminus \sigma_{R'}} [b_R] - b_{R'}\;, \qquad \forall R',R : R'\subset R\;;\nonumber
\end{eqnarray}
where $\Tr_{\setminus}$ is a trace over all variables excluding those listed in the subscript.
The beliefs recovered for a correct region selection are exactly the marginal probabilities, and are consistent with those determined by linear response $b_R=b^{LR}_R$.

The regions prescribed by a junction tree have a maximum size that depends on the graph width, if this topological property is small then CVM can be an efficient way to calculate the free energy. If the graph is uncorrelated ($J=0$) then the NMF approximation is exact, with $c_R=1$ for single variable regions and $c_R=0$ for all other regions. For purposes of evaluating the energy (\ref{eq:energyCVM}) we take $b_s=\prod_{i\in s} b_i$ in NMF. If the graph is a tree (or forest) the Bethe approximation is exact, with edge regions for every element in $I$, in addition to all vertex regions. In the Bethe case: $c_s=1\;, \forall s\in I$;  $c_i=1 - \sum_{R \subset I: i \in R} c_R$; and $c_R=0$ otherwise. 

Unfortunately, junction trees are also impractical in general. Typically the width of the graph is large, requiring large regions for an exact solution, so that the evaluation of the entropy (\ref{eq:entropyCVM}) is impractical. One is therefore interested in approximations; fortunately NMF and Bethe are found to be good, or asymptotically exact in many circumstances. Given that we resort to these methods in cases where the method is not exact, how should one construct the marginal probabilities? Two options exist for the regions exploited in the approximation: one can use the maximum entropy estimate $P_R\approx b_R$; or one can use the linear response estimtate $P_R \approx b_R^{LR}$ about the minima of $F_{CVM}$. In general these estimates differ; one measure of the quality of the variational approach is the amount of agreement between these values. This paper discusses a modification to $F_{CVM}$ that allows for exact agreement.

Let us parameterize the beliefs over single variables in a manner comparable to (\ref{eq:linearresponsemarginals}), using the set of magnetization ($C_i$), and in the case of Bethe symmetric pair correlation parameters ($C_{ij}=C_{ji}$) 
\begin{equation}
  b_i = \frac{(1 + C_i \sigma_i)}{2}\;; b_{ij} = b_i b_j + \frac{C_{ij}\sigma_i \sigma_j}{4}\;,
\end{equation}
where $C_{ij}=0$ for NMF. By this method the constraints (\ref{eq:constraint}) are made redundant and we have an unconstrained minimization problem, subject to a parameter range $0 < b_R < 1$. We exclude the possibility of boundary values $\{0,1\}$ since to apply linear response we will assume a minima in which all parameters can fluctuate.

In this paper we consider the following modification to the entropy approximation: we introduce, in the case of NMF the constraint that the self-response and magnetization agree as per the exact free energy 
\begin{equation}
  \chi_{ii} = 1 - C_i^2\;,\; \qquad \forall i\;. \label{eq:diagonalconstraint}
\end{equation}
We can introduce a Lagrange multiplier in the standard form to write the entropy approximation for NMF 
\begin{equation}
  S^{N}_{\lambda} = -\sum_i \Tr\left[b_i \log b_i\right] - \sum_i \lambda_i  \left( \left(1 - C_i^2\right) - \chi_{ii}\right)/2\;.
\end{equation}

Within the Bethe approximation agreement between $b_{ij}=b^{LR}_{ij}$ requires the additional constraint
\begin{equation}
  \chi_{ij} = C_{ij}\;;\qquad \forall ij \in I \;,\label{eq:offdiagonalconstraint}
\end{equation}
the entropy approximation for Bethe becomes
\begin{equation}
  S^{B}_{\lambda} = S^{N}_{\lambda} - \sum_{ij \in I} \left\lbrace \Tr \left[b_{ij} \log \left(\frac{b_{ij}}{b_ib_j}\right)\right] - \lambda_{ij}  \left( C_{ij} - \chi_{ij}\right)\right\rbrace\;, 
\end{equation}
The entropies presented are a generalization of those discussed in ~\cite{Raymond:MFM}, where only the off-diagonal constraints were considered ($\lambda_i=0,\; \forall i$).
 
\subsection{Saddle-point equations and reaction terms}
A minima of the free energy requires that the derivatives with respect to the variational parameters are zero. The derivative with respect to $C_i$ is 
\begin{equation}
  0 = \atanh(C_i) - H_i - \sum_{j (\neq i)}J_{ij}C_j + L_i - \lambda_i C_i \label{eq:dFdCi}\;,
\end{equation}
where $L_i=L^N_i=0$ for NMF, and for Bethe
\begin{equation}
  L^B_i = \sum_{j} c_{ij} \Tr\left[b_j \frac{\sigma_i}{2}\log\left(\frac{b_{ij}}{b_i}\right) \right]\;.
\end{equation}
For the Bethe method we must also consider the derivative with respect to $C_{ij}$
\begin{equation}
  0 = L^B_{ij} - J_{ij} = \Tr \left[\frac{\sigma_i \sigma_j}{4}\log b_{ij}\right] - J_{ij} + \lambda_{ij} \label{eq:dFdCij}\;.
\end{equation}

Note that both the entropic term $\lambda_i C_i$ and the leading order of $L_i$, are reaction terms (proportional to $C_i$); for fully connected models these reaction terms are well understood in the large system limit~\cite{Opper:TAPMpre}. When $\lambda$ is taken to be non-zero and fixed by linear response, we find that quite generally we recover the Onsager reaction term, as later discussed. When the constraint (\ref{eq:diagonalconstraint}) is not enforced $\lambda_i=0$, in applying the Bethe approximation the Onsager reaction term is recovered for the case of independent identically distributed couplings.
 
When considering variation of the free energy we will treat both $\chi$ and $\lambda$ as fixed external parameters, the variation is restricted to $C$.
The Hessian, with components
\begin{equation}
  Q_{s,s'} = \frac{\partial^2 F_{CVM}(b)}{\partial C_s \partial C_{s'}}\;,
\end{equation}
is required to be positive definite at the minima.

Supposing $C=C^*$ describes the minimizing arguments for $\{H,J\}$, in response to a small variation in the fields $H+\delta H$ the new minima $C=C^*+\delta C$ can be determined from the quadratic order expansion of the free energy
\begin{eqnarray}
  F^X_{\lambda}(H+\delta H, J) &=& \min_{\delta C} \biggr\lbrace F^X_{\lambda}(H, J) + \delta C Q^{X} \delta C/2 \nonumber\\
	&-& \sum_i \delta_{iz} \delta H_z (C_i + \delta C_i) \biggr\rbrace\;, \label{eq:Quadratic}
\end{eqnarray}
for NMF ($X=N$) or Bethe ($X=B$); this will be the basis for constructing the linear response identities.

\subsubsection{NMF}
\label{sec:NMFHess}

In the case of NMF the components of the Hessian are
\begin{equation}
  Q^{N}_{i,j}  = -J_{i,j} + \delta_{i,j} [\frac{1}{1-C_i^2} - \lambda_i] \;.
\end{equation}
The argmin for NMF is
\begin{equation}
  \delta C_i = \sum_z [(Q^N)^{-1}]_{iz} \delta H_z\;.
\end{equation}
Now, the linear response identity in the limit $\delta H \rightarrow 0$ allows us to decompose $\delta C$ as a sum of perturbations on individual fields at leading order
\begin{equation}
  \delta C_i = \sum_z \chi_{iz}\delta H_z\;.
\end{equation}
As such
\begin{equation}
  \chi_{i,j} = [(Q^N)^{-1}]_{i,j} = [(\Phi^N - J)^{-1}]_{i,j}\;.
\end{equation}
We introduce notation $\Phi^N$ to denote the entropic (approximate) part of $Q$, separated from the energetic (exact) part. 
This matrix gives us estimates for all pair correlations, as well as those within regions. The constraint (\ref{eq:diagonalconstraint}) is
\begin{equation}
  [\chi^{-1}]_{i,i} = \frac{1}{1-C_i^2} - \lambda_i \label{eq:DiagonalIdentityNMF}\;.
\end{equation}

We have a closed set of equations for $\{C,\lambda\}$ in (\ref{eq:dFdCi}) and (\ref{eq:DiagonalIdentityNMF}). Interestingly these equations are exactly those that define the adaptive TAP. It was shown by Opper and Winther that a solution to the above equations does indeed reproduce the Onsager reaction term for correlated and uncorrelated distributions of $J$ (and without prior knowledge of the statistics)~\cite{Opper:TAPMpre}. 

\subsubsection{Bethe}

To describe the Hessian at the Bethe level variation of the pair correlation parameters must be considered
\begin{eqnarray}
  \!Q^B_{i,j}\!\! &=&\!\! Q^{(1)} \!= \!Q^N_{i,j} \!+\! c_{ij} \left\lbrace\Tr \left[\frac{\sigma_i\sigma_j}{4}\left(\log b_{ij} + \frac{b_i b_j}{b_{ij}}\right) \right]\right\rbrace \nonumber \;;\\
  \!Q^B_{ij,k}\!\! &=&\!\! Q^{(2,1)}_{ij,k} = c_{ij}\left\lbrace\delta_{i,k} \Tr \left[\frac{b_{j}\sigma_j}{8 b_{ij}}\right] + \delta_{j,k}\Tr \left[ \frac{b_{i}\sigma_i}{8 b_{ij}}\right]\right\rbrace \nonumber \;;\\
  \!Q^B_{ij,kl}\!\! &=&\!\! Q^{(2)}_{ij,ik} = c_{ij}\delta_{ij,kl}\Tr \left[ \frac{1}{16 b_{ij}}\right] \;.
\end{eqnarray}
If we break the vector $\delta C$ into two parts, one describing single variable variation $\delta C^1 = \{\delta C_i\}$ and one describing pairwise variation $\delta C^{2} = \{\delta C_{ij}\}$ we can determine the minimizing arguments of (\ref{eq:Quadratic}) from 
\begin{eqnarray}
  0 &=& Q^{(2,1)}\delta C^1 + Q^{(2)} \delta C^2 \;;	\\
  \delta H &=& Q^{(1)}\delta C^1 + {[Q^{(2,1)}]}^T \delta C^2\;.
\end{eqnarray} 
Eliminating $\delta C^2$ we can proceed as for NMF
\begin{eqnarray}
  \chi = [Q^{(1)} - [Q^{(2,1)}]^T {[Q^{(2)}]}^{-1}Q^{(2,1)}]^{-1} = {[\Phi^B - J ]}^{-1} \;.
\end{eqnarray}

Calculating the off-diagonal component of $\Phi_{ij}$ where $(i,j) \in I$, we find
\begin{equation}
  [\chi^{-1}]_{i,j} \!=\! \Phi^B_{ij} - J_{ij} \label{eq:SessakMonasson}\;.
\end{equation}
where the expressions for $i\neq j$ and $i=j$ are
\begin{eqnarray}
\!\!  \Phi^B_{ij} \!\!\!&=&\!\!\!\!\! c_{ij} \left[\Tr [\frac{\sigma_i \sigma_j}{4} \log b_{ij}] \!-\! \frac{C_{ij}}{(1\!-\!C_i^2)(1\!-\!C_j^2)\!-\!C_{ij}^2}\right],\\
\!\!  \Phi^B_{ii} \!\!\!&=&\!\!\!\!\! \frac{1}{1 \!-\! C_i^2}\!\left[1 \!+\! \sum_{j (\neq i)}c_{ij}\frac{C_{ij}^2}{(1 - C_i^2) (1 - C_j^2) - C_{ij}^2} \right].
\end{eqnarray}
Applying the constrained method with (\ref{eq:offdiagonalconstraint}), we find that for $i\neq j$ in (\ref{eq:SessakMonasson}) we recover the Sessak-Monasson expression for calculation of correlations~\cite{Sessak:SME}.

\section{Implementation}
\subsection{Weak coupling limit}

When couplings $J_{ij}$ (and consequently $C_{ij}$) are small in absolute value the solution to the free energy can be found as an expansion about the decoupled case. Using the exact entropy terms  $L^{*}$ and $\Phi^{*}$ one can determine as expansions in $\{J\}$ the corresponding solutions $C^*$,$\chi^*$ and $\lambda^*$ ($\lambda^*=0$ in the exact case). 

For $J=0$ Bethe and NMF are exact. We can make an expansion of our expressions assuming $C=C^*+\delta C$ and $\lambda=\lambda^*+\delta \lambda$, with both $\delta C$ and $\delta \lambda$ small. We linearize our equations, all terms are evaluated for the exact solution $\{C^*,\lambda^*,\chi^*\}$ in the following expressions. From (\ref{eq:dFdCi}) 
\begin{equation}
  % 
%  0 = Q^N_{i,i} \delta C_i +\sum_{j (\neq i)}Q_{i,ij}\delta C_{ij} + ({L^X_i}^* - {L_i^{exact}}^*) - \lambda_i C_i^* \;,
  0 = \sum_{s: i\in s} Q^X_{i,s} \delta C_s + ({L^X_i} - L_i^*) - \delta \lambda_i C_i \;.  \label{eq:linearizeddFdM}
\end{equation}
For the Bethe method we require, from (\ref{eq:dFdCij}),
\begin{eqnarray}
  0 &=& (L^B_{ij} - L_{ij}^*) + \delta \lambda_{ij} + \sum_{s = i,j,ij} Q_{ij,s}\delta C_{s}\;, \label{eq:linearizeddFdC}
\end{eqnarray}
which in the constrained case (\ref{eq:offdiagonalconstraint}) defines $\delta \lambda_{ij}$. Since $\delta \lambda_{ij}$ enters no other equations it does not influence the error on $C$.

We can expand $\Phi$ about the exact result, distinguishing the on and off-diagonal terms
\begin{equation}
  %X = \chi^0(C^*)[Diag[\frac{\Phi^N_{ii}}{\partial C_i} \Delta C_i] + Diag[\lambda_i] + [\Phi^X(C^*)-\Phi^{exact}(C^*)]]\chi^0(C^*)\\
%  {\tilde \Phi} = [Diag[\frac{-2 M_i}{(1-M_i^2)^2} \Delta C_i] + Diag[\lambda_i] + [\Phi^X-\Phi^{exact}]]
  {\tilde \Phi}^X_{i,j} = (\Phi_{i,j}^X - \Phi_{i,j}^*) - \delta_{i,j}\delta \lambda_i + \sum_{s}\frac{\partial \Phi^X_{i,j}}{\partial C_s} \delta C_s 
\end{equation}
The constraints (\ref{eq:diagonalconstraint}) and (\ref{eq:offdiagonalconstraint}) dictate respectively
\begin{equation}
  % (1- m^2)_ii = [Exact + Phi^X]^{-1}_{ii} = [ Exact + Phi^X^* + (Phi^X-Phi^X^*)]^{-1}_{ii}
  % (1- m^2)_ii = [Exact + Phi^X]^{-1} = [ Exact + Phi^* + (Phi^X^* - Phi^*) + (Phi^X-Phi^X^*)]^{-1}_{ii}
  % A = [M + dM]^{-1}_{ii}
  % A = M^-1_{ii} - {M [dM] M}_{ii}
  % dM = (Phi^X^* - Phi^*) + (Phi^X-Phi^X^*)
   -2 C_i \delta C_i = - [\chi {\tilde \Phi}^X \chi]_{ii} \;; \qquad \delta \chi_{ij} = - [\chi {\tilde \Phi}^X \chi]_{ij} 
\end{equation}
where $\delta \chi_{ij}=\delta C_{ij}$ in the latter case, otherwise $\{\lambda_{i}=0\}$ or $\{\lambda_{ij}=0,\delta \chi_{ij}\neq \delta C_{ij}\}$ if the respective constraints are not introduced.

% LNMF - Lexact = - t_i (1-t_j^2) J_{ij}^2
% LBethe - Lexact = + 2 t_i (1-t_j^2)(1-t_k^2) j_{ij} j_{ik} J_{jk}
% PhiiiNMF -PhiiiExact = - J_{ij}^2 (1-t_j^2)
% PhiiiBethe - PhiiiExact = + 2(1-t_j^2)(1-t_k^2) j_{ij} j_{ik} J_{jk}
% PhiijNMF - PhiijNMF = 2 Jij^2 t_i t_j + 2/3 J_{ij}^3
% 
The errors in responses $L_i$, as well as on and off-diagonal elements of $\Phi$ are significant in determining the errors.
At leading order ($\doteq$) in the weak coupling limit ($J$ small)
\begin{eqnarray}
  %\frac{1}{1-M_i^2}\sum_j c_{ij} \frac{C_{ij}^2}{(1-M_i^2)(1-M_j^2)-C_{ij}^2} where Cij->(1-Mi^2)(1-Mj^2) J_{ij}
  %Delta Phi B = -2 Jij^2 m1 m2 
  \Phi^N_{i<j} - \Phi^{exact}_{i<j} &\doteq& 2 J_{ij}^2 t_i t_j + \frac{2 J_{ij}^3}{3}  \label{eq:PhiNlinearized}\\
  \Phi^B_{i<j} - \Phi^{exact}_{i<j} &\doteq& \sum_{k (\neq i,j)} - 2 J_{ik} J_{jk} T_k \biggr(J_{ijk} \nonumber \\
  &+& 2 \left(J_{ikj} +  J_{jki}\right)\biggr) - 2 J_{ik}^2 J_{jk}^2J_{jk} \label{eq:PhiBlinearized}
\end{eqnarray}
where $t_i=\tanh(J_i)$, $T_i=1-t_i^2$ and we define $J_{ijk}= J_{ik}J_{jk}t_i t_jT_k$; and 
\begin{equation}
  L^X_i -L^*_i \doteq t_i D^X_i  \;;\;\qquad \Phi^X_{ii} - \Phi^*_{ii} \doteq D^X_i \;. \label{eq:cons1}
\end{equation}
For the NMF and Bethe methods we define respectively
\begin{equation}
  D_i^N = -\sum_{j (\neq i)} J_{ij}^2T_j\;;\qquad
  D_i^B = 2 \sum_{j<k (\neq i)} J_{jk}J_{ij}J_{ik}T_j T_k\;.\nonumber
\end{equation}
In the Bethe method where $C_{ij}\neq \chi_{ij}$ the error on $C_{ij}$ is dominated by
\begin{equation}
  L^B_{ij}- L^*_{ij} \doteq \sum_{k (\neq i,j)}J_{ik}J_{jk} T_k\;.\label{eq:dC2}
\end{equation}
In (\ref{eq:PhiNlinearized}) and (\ref{eq:PhiBlinearized}) we demonstrate also the leading order diagram relevant for high temperature at $O(\beta^3)$ and $O(\beta^5)$ respectively. 

%The various sources of errors act in combination, with different terms being of greater or lesser significance depending on the constraints implemented:
%At leading order
%\begin{eqnarray}
%  \frac{\partial \Phi^X_{ii}}{\partial C_i} &=& \frac{2 t_i}{(1-t_i^2)^2}  \;,%+ \sum_j c_{ij} 2 t_i(1 + 3 t_j^2) J_{ij}^2
%\end{eqnarray}
%and the additional terms relevant to Bethe are at leading order
%\begin{eqnarray}
  %%(4 (m1 m2) c)/((-1 + m1)^2 (1 + m1)^2 (-1 + m2)^2 (1 + m2)^2) + ((-2 + 6 m2^2 + 6 m1^2 (1 + m2^2)) c^2)/((-1 + m1)^3 (1 +     m1)^3 (-1 + m2)^3 (1 + m2)^3) +
%  \frac{\partial \Phi^B_{ii}}{\partial C_{j}} &=& 2 t_j J_{ij}^2 + O(t J^3) \\ %\frac{1}{1-M_i^2}\sum_j c_{ij} \frac{C_{ij}^2}{(1-M_i^2)(1-M_j^2)-C_{ij}^2} = \frac{1}{1-M_i^2}c_{ij} \frac{C_{ij}^2 (1-M_i^2)2 M_j}{((1-M_i^2)(1-M_j^2)-C_{ij}^2)^2} = c_{ij} \frac{J_{ij}^2 2 t_j}{(1-J_{ij}^2(1-t_i^2)1(1-t_j^2))^2} = J_{ij}^2
%  \frac{\partial \Phi^B_{ij}}{\partial C_{ij}} &=& 4 t_i t_j J_{ij} - 2 J_{ij}^2 + O(tJ^2,J^3) \\
%  \frac{\partial \Phi^B_{ii}}{\partial C_{ij}} &=& 2(1 - t_i^2) J_{ij} + O(J^3) \\ 
%  \frac{\partial \Phi^B_{ij}}{\partial C_i} &=& 2 t_j(1 + 3 t_i^2) J_{ij}^2 + O(t_i J^3)\;.
%\end{eqnarray}
%and we know at leading $\Phi^X = \mathrm{Diag}[1/(1-t_i^2)-\lambda_i]-J$.
%We can solve the system of linear equations to determine leading order errors.

We calculate errors for both the weak coupling (small $J$) and high temperature ($J$ and $H$ are $O(\beta)$) cases solving the linearized equations. We summarise the consequences for the error in $C_i$, $\chi_{i \neq j}$ and $C_{ij}$ according to constraints introduced (left label in list). For NMF errors are
\begin{itemize}
  \item[$\emptyset$] From (\ref{eq:linearizeddFdM}) and (\ref{eq:cons1}) $\delta C_i$ is determined as $O(J^2,\beta^3)$. The response error $\delta \chi_{ij}$ is $O(J^2,\beta^3)$.
  \item[(\ref{eq:diagonalconstraint})] We find $\lambda^N_i=D^N_i$, removing the most significant source of error in $\delta C_i$, the error on the magnetization improves to $O(J^3,\beta^4)$, the error on $\delta \chi_{ij}$ remains limited to $O(J^2,\beta^3)$ by the error (\ref{eq:PhiNlinearized}).
\end{itemize}%NB ALWAYS DOMINATED BY delta Phi Offdiag ~ delta Cij ~  delta Phi Ondiag * J 
For Bethe errors are
\begin{itemize}
  \item[$\emptyset$] $\delta C_i$ and $\delta \chi_{ij}$ are $O(J^3,\beta^4)$.
  \item[(\ref{eq:diagonalconstraint})] We find $\delta C_i$ is improved to $O(J^4,\beta^5)$, $\delta \chi_{ij}$ remains $O(J^3,\beta^4)$. Error sources (\ref{eq:cons1}) are improved, but (\ref{eq:dC2}) remains a significant constraint on accuracy of $\delta \chi_{ij}$.
  \item[(\ref{eq:offdiagonalconstraint})] $\delta C_i$ remains $O(J^3,\beta^4)$ but $\delta \chi_{ij}$ is improved to $O(J^4,\beta^4)$. The errors on $\delta C$ are made independent of (\ref{eq:dC2}), but the error sources (\ref{eq:cons1}) are unimproved. 
  \item[(\ref{eq:diagonalconstraint},\ref{eq:offdiagonalconstraint})] The combined effect is to remove the most significant sources of error, both $\delta C_i$ and $\delta \chi_{ij}$ become $O(J^4,\beta^5)$. The remaining error on $\delta \chi_{ij}$ is limited at leading order only by (\ref{eq:PhiBlinearized}).
\end{itemize}%NB ALWAYS DOMINATED BY delta Phi Offdiag ~ delta Cij ~  delta Phi Ondiag * J 
For Bethe introducing the constraint (\ref{eq:offdiagonalconstraint}) always reduces the error on $\delta C_{ij}$, which is $O(J^2,\beta^2)$ in the standard method. % (\ref{eq:dC2})

\subsection{Iterative scheme}
The non-convex nature of the constraints we are introducing makes algorithm development a challenge, but we can solve in general these equations for weak-coupling, with a naive iterative scheme
\begin{equation}
  C_i^{t+1} = \tanh\left(H_i + \sum_j J_{ij}m_j + \lambda^t_i C_i^t - L^t_{i}\right)\;.\label{eq:iteratemag}
\end{equation}
If applying the constraint (\ref{eq:diagonalconstraint}), we can simultaneously infer
\begin{equation}	 	 
  \lambda^{t+1}_i = \lambda^t_i - \Phi^t_{ii}((1-(C^t_i)^2) - \chi^t_{ii})\Phi^t_{ii} \;,\label{eq:iteratelambdaondiag}
\end{equation}
otherwise $\lambda_i=0$. Applying constraint (\ref{eq:offdiagonalconstraint}), for the Bethe method, 
\begin{equation}
  \chi_{ij}^{t} = [(\Phi^t-\beta J)^{-1}]_{ij}\;;\qquad C_{ij}^{t+1}=\chi_{ij}^{t}\;.\label{eq:iterateCijLR}
\end{equation}
with $\lambda_{ij}$ fixed by (\ref{eq:dFdCij}); otherwise $\lambda_{ij}=0$ and we fix 
\begin{equation}
b_{ij}^{t}=\argmin_{b_{ij}} \{b_{ij}\log b_{ij} - J_{ij}\Tr[ b_{ij}\sigma_i \sigma_j] : C^t_i,C^t_j \} \;.\label{eq:iterateCijMaxEnt}
\end{equation}
 To fix $b^t_{ij}$ at fixed $C^t_{i}$ and $C^t_j$ is equivalent to fixing $C_{ij}^t$. 

The instantaneous mean field is used to update the magnetization in (\ref{eq:iteratemag}), a linear expansion of (\ref{eq:diagonalconstraint}) is used to determine (\ref{eq:iteratelambdaondiag}), a naive iteration matching successively the linear responses is used in (\ref{eq:iterateCijLR}). At large $|J|$ (\ref{eq:iteratemag})-(\ref{eq:iterateCijLR}) can be unstable individually or in combination, damping and annealing can be effective strategies to arrive at a solution for strong coupling. The procedure (\ref{eq:iterateCijMaxEnt}) is one of convex optimization and doesn't contribute to instability.

\subsection{Strong coupling regime experiments}
\begin{figure}[!t]
\centering
\includegraphics[width=3.5in]{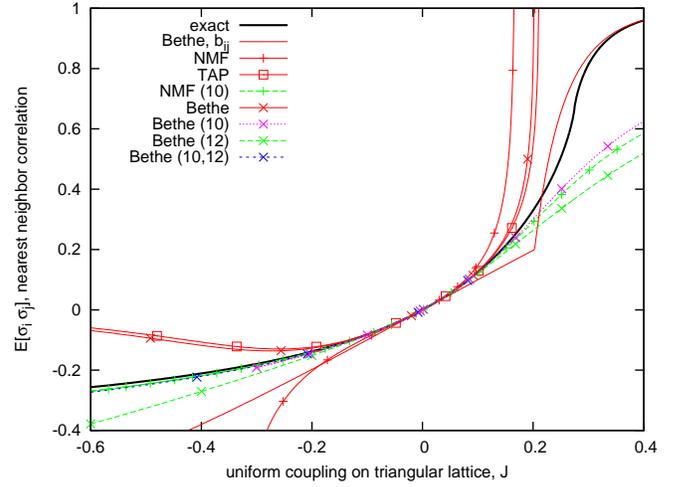}
\caption{\label{figure:energy} Full correlations estimates on nearest neighbors based on linear response $b^{LR}_{ij}$, compared to the exact result (black think curve), and the parameters $b_{ij}$ determined for a standadrd implentation of the Bethe approximation (thin red). Curves are labeled in the legend according to the constraints introduced. For negative $J$ the new methods perform admirably compared to standard implementations. All methods perform poorly in the vicinity of the phase transition, the paramagnetic solutions of the new methods can be stable even beyond the true critical point $J>0.275$, though performance is poor. }
\end{figure}
%\begin{figure}[!t]
%\centering
%\includegraphics[width=\linewidth]{FIGURES/figure2_5by5_netstat}
%\caption{\label{figure:Cnnn}Next nearest neighbor correlation for homogeneous model $J_{ij}=J$, $H_i=0$ on $L=5$ triangular% lattice with periodic boundary conditions. }
%\end{figure}
We consider a simple model the triangular lattice model with uniform couplings $J_{ij}=J$ and zero fields $H_i=0$ in the large system limit. This model is problematic for standard Bethe and NMF for several reasons: it involves short loops not accounted for by the region selection; there is a continuous symmetry breaking transition at $J=0.275$ with associated long range correlations~\cite{Baxter:ESM}; for $J<0$ there is frustration; for $J<0$ there are Kosterlitz-Thouless transitions, but no symmetry breaking transitions~\cite{Wannier:A,Stephenson:IMSC}. For these reasons Bethe and NMF estimates for $b_R$ or $b_R^{LR}$ can be poor. The solution can be found for our new methods by Fourier analysis. Figure \ref{figure:energy} presents a comparison of methods. We present only the solution found continuously from $J=0$ by the iterative method, and we do not present the symmetry breaking solutions at $J<0$, where they exist. 

For $J>0$ the paramagnetic ($\{C_i=0\}$) solutions are, for small $|J|$, in close agreement with the exact result. Amongst new methods all but the doubly constrained approximation (Bethe with (\ref{eq:diagonalconstraint}) and (\ref{eq:offdiagonalconstraint})) remain locally stable well beyond the true ferromagnetic transition point. The large $J$ paramagnetic solutions do have some problems: the correlation estimate $E[\sigma_i\sigma_j]$ becomes poor compared to the standard implementations that undergo symmetry breaking transitions; negative entropy can be found (when $C_{ij}>0.652$); and finally the iterative method struggles to converge without strong damping. The doubly contrained method on the other hand has a Hessian that becomes singular (for the paramagnetic solution) at $J=0.16$, in advance of even the standard NMF instability $J=1/6$; with simple iterative methods we are unable to construct a magnetized solution.

For $J<0$, the frustrated regime, performance of the new methods is a clear improvement over standard implementations. We find a similar pattern of results with respect to other components of $\chi$ (longer range correlations), and in testing lattices of finite size. An instability towards symmetry breaking on the tripartite sublattices causes the termination of the line Bethe (\ref{eq:diagonalconstraint}) for $J<0$; a similar instability affects the standard NMF implementation.  

\section{Conclusion}
The fundamental objects of the cluster variational methods are the beliefs, which are determined by maximum entropy. One would expect that these beliefs would be reproducible by linear response up to some small error, but for the standard method this is not true. The discrepancy provides information on global graph structure that we have exploited in the new approximation where max entropy and LR are made consistent. In this article we have required consistency with the quadratic order LR relations, which we expect are of greater importance than those of higher order. 

As part of our method we recover the Sessak-Monasson equation for pair correlations~\cite{Sessak:SME}. Furthermore, for the case of spins and working from the NMF framework we derive the adaptive-TAP equations~\cite{Opper:TAPMpre}. The arguments by which these well known equations were originally derived is very different from our own, offering a point of comparison. In our derivation we note that we arrive at these results using simple region selection rules in the cluster variational framework, without assuming weak correlations; in the case of adaptive TAP and Sessak-Monasson we can improve performance by moving to higher order region approximations~\cite{Raymond:MFM}.

Our framework is also very flexible with respect to the nature of the Hamiltonian; the result is easily generalized to other pairwise models, or with small modifications to multi-body interactions. We can also hope that some of the powerful expansion and algorithmic methods mentioned in the introduction will be made compatible. Application to the Inverse problem are also promising, and relatively simple compared to standard implementations of the CVM method~\cite{Raymond:MFM,RicciTersenghi:InverseIsing}. 


%\section*{Acknowledgment}




\bibliographystyle{IEEEtran}
%\bibliography{../Bibliography}
\vspace{-0.1cm}


%\begin{thebibliography}{1}

%\end{thebibliography}

\end{document}